\documentclass[11pt]{article}

\usepackage{amsmath}    % need for subequations
\usepackage{amsfonts}
\usepackage{graphicx}   % need for figures
\usepackage{placeins}
\usepackage{verbatim}   % useful for program listings
\usepackage{color}      % use if color is used in text
\usepackage{caption}
\usepackage{subcaption}
\usepackage{hyperref}   % use for hypertext links, including those to external documents and URLs
\usepackage{enumitem}
\usepackage[font=small,skip=0.1pt]{caption}

\begin{document}

\noindent
{\large \bf Evaluation of Coded Aperture Radiation Detectors using a Bayesian Approach}\\
\noindent
\small
{K. ~Miller$^{*}$, P. ~Huggins$^*$, A. ~Dubrawski$^*$, S. ~Labov$^{**}$, K. ~Nelson$^{**}$} \\ 
\noindent
{${}^*$ Auton Lab, Carnegie Mellon University} \,\, 
{${}^{**}$ Lawrence Livermore National Laboratory} 

%\maketitle
\normalsize
\vskip 0.075in
\noindent {\bf Introduction.} We investigate the utility of coded aperture (CA) for roadside radiation threat detection applications.  With coded aperture, information in the form of photon quantity is traded for directional information. Whether and in what scenarios this trade-off is beneficial is the focus of this study. 
We quantify the impact of a masking approach by comparing performance with an unmasked approach in terms of both detection and localization of a roadside nuclear threat. 
% In this problem, a mobile detection vehicle drives along roadways in an effort to detect and locate threat objects. 
We simulate many instances of a drive-by scenario via Monte Carlo using empirical observations from the RadMAP project {[}1{]} to obtain background photons and synthetic injection of threat source. 
Simulation results suggest that at 0.1\%false positive rate (FPR) the masked detector suffers 20-50\% loss of detection probability for weak sources, but only 1-1.5\% for moderate source intensities. 
The masked approach also demonstrates consistent improvement in localization performance ranging from 0.7-2.2m across all source intensities investigated.
 
\vskip 0.05in
\noindent {\bf Methods.} We reduce empirical background data to gamma observations aggregated to one second intervals. From these data, a generative model of background radiation is learned. Simulation replicates are then conducted by randomly drawing source location and background measurements and then injecting in them additional photon counts due to the source. Observations are scored using a pseudoinverse decoding method or censored energy windowing ({CEW}) for the masked and unmasked detectors, respectively. Scores are aggregated using either {Bayesian} aggregation ({BA}) or weighted combining ({WC}) method to produce score distributions over two-dimensional spatial locations. Finally, detection and localization are validated by reporting the value and location corresponding to the maximal score over a neighborhood of $\pm 10$m centered on the true source location. We compare the threat detection performance of the detectors and associated scoring algorithms under these two approaches to observation aggregation.

The generative model of background is learned as a vector of energy bin specific rates for the location of each empirical observation. Location specific rates are taken as the maximum a posteriori estimates of the empirical data within a 20m radius using a Poisson model of the data and a Gaussian prior on the rates. The Gaussian prior is taken to the have mean and covariance estimated over all empirical observations. This approach can be thought of as a learning a moving average of background rates over locations, yielding a generative model of background that preserves systematic variation over space. 

Data observed using the CA is modeled as a linear combination of uniform (angle independent) background and exposure to source for each detector element. Such linear combinations from each of the 100 detector elements are combined into an exposure matrix with a column of ones and a column of exposure coefficients, specific to a hypothesized angle to source. The background and source strengths that maximize the likelihood of an observation under a Gaussian model with Gaussian priors (uninformative for the source) can be found using the Moore-Penrose inverse of the exposure matrix augmented with two additional rows for the priors. We call this the psuedoinverse decoding ({PID}) method and take the predicted source strength as the threat score to be consumed by the data fusion routine. {CEW} aggregates across individual detector elements and predicts source strength as the residual from a linear model that estimates within window counts from remaining out-of-window energy bins. A score is then calculated as the SNR; residual over the square root of predicted counts.

BA operates under a known source spectrum and intensity. The likelihood ratios of observing each score under an assumed source spectrum, intensity, and location, with and without source present, are aggregated across all observations. For these simulations, source spectrum is assumed known and was chosen as an artificial source template created in such a way that it is representative of shielded special nuclear material (SNM) under the assumption of a single low energy peak. We marginalize hypothetical source intensities. The WC method computes an aggregate SNR score by summing source estimates over exposure and dividing by the square root of the sum of background estimates over squared exposure. Exposure is a function of the hypothetical source location and thus WC yields a spatial distribution of scores.

\vskip 0.05in
\noindent
{\bf Experiments.} We ran 12,000 replicate Monte Carlo simulations for source intensities uniformly distributed within 6 bands ranging from 1$\mu$Ci to 750$\mu$Ci. Background data were taken from four hours of RadMap collection using a 10 by 10 NaI detector array, over 4 hours in an urban area of California. Total background detected in empirical observations was approximately 10,000 counts per second. NaI sensor data were processed by first removing detection events with energy levels less than 30keV, and converting to spectra with a 1s live time and 128 quadratically spaced bins. Injected sources were placed up to 40m from the roadway with uniform distribution over location. This produced a stand-off distance from 1 to 40m with an average of 22m.

Figure~1 shows Receiver Operating Characteristic (ROC) curves for the four combinations of detector and aggregation approach for three selected bands of source strengths. Using BA, the masked detector performance is 50-80\% of that of the unmasked detector at 0.1\% FPR for source intensities of 75$\mu$Ci and below. However, the detection performance loss is only 1-1.5\% for source intensities between 100 and 750$\mu$Ci. Figure~2 shows box plots of localization error for detections made at the 0.1\% FPR threshold. Outliers have been removed for clarity. Table~1 shows the probability of detection at the 0.1\% FPR threshold for each method. 

\begin{figure}
\centering
\begin{subfigure}[b]{0.24\textwidth}
   \includegraphics[height=1.75in]{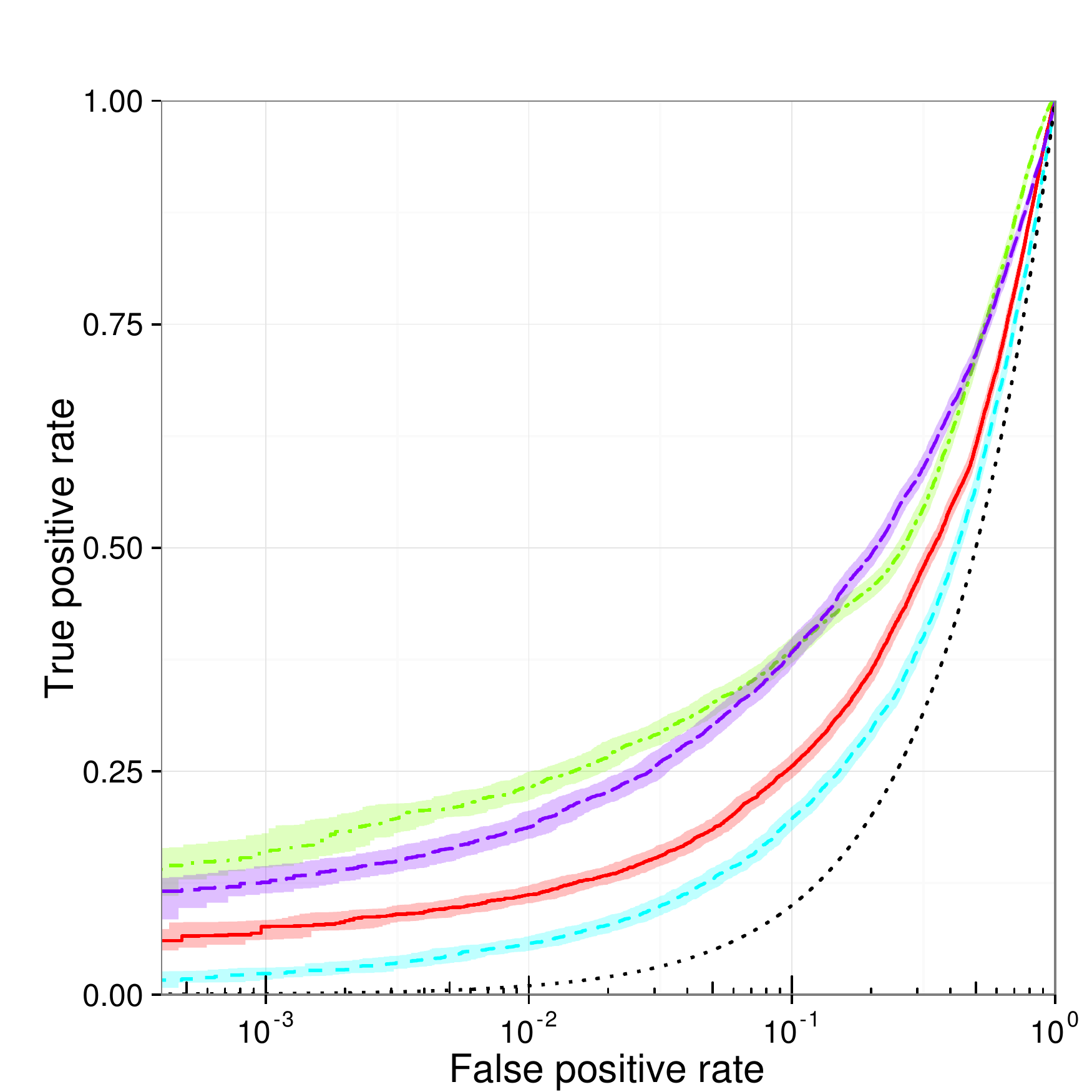}
   \caption{$5$-$7.5\mu$Ci}
   \label{ROC1}
\end{subfigure}
\begin{subfigure}[b]{0.24\textwidth}
   \includegraphics[height=1.75in]{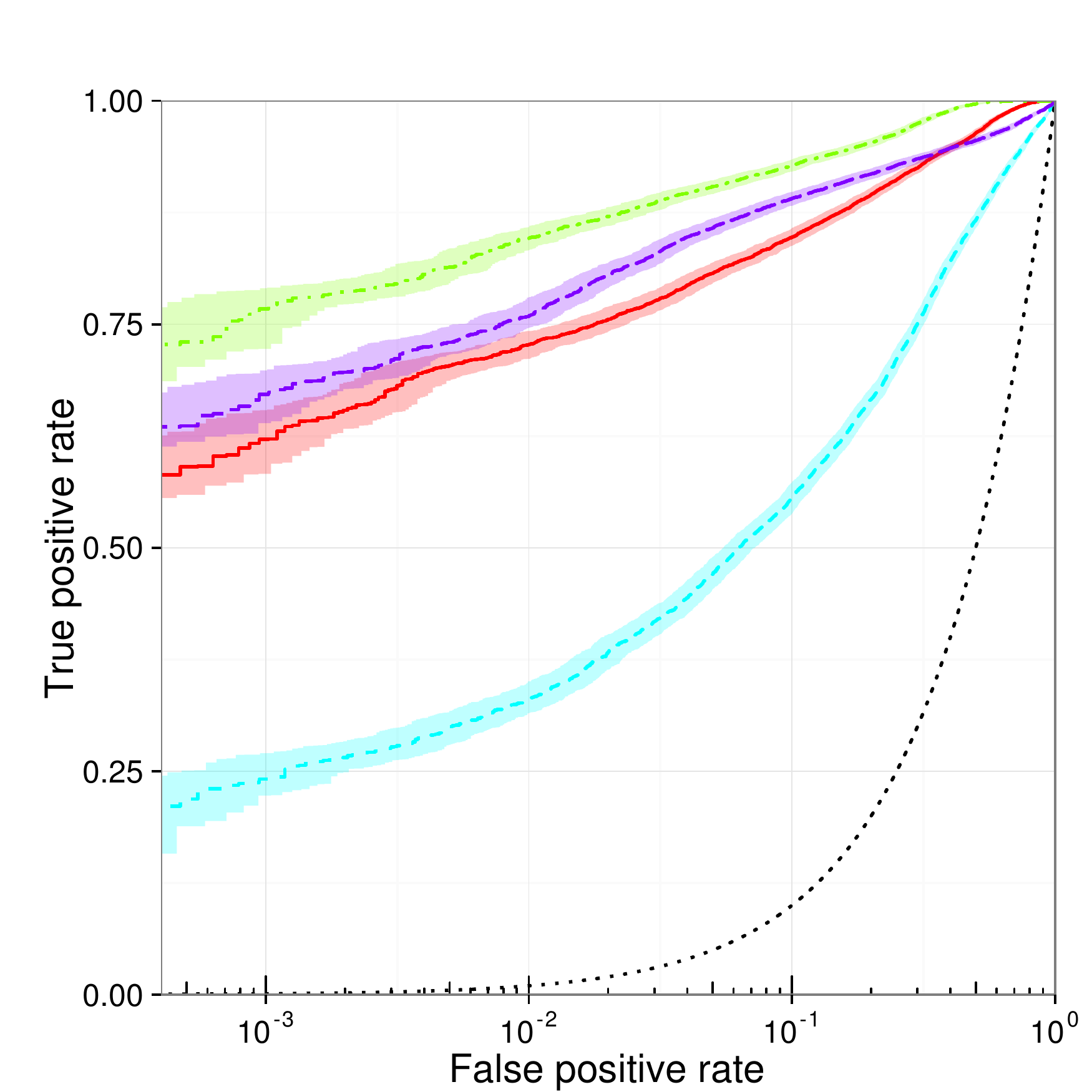}
   \caption{$50$-$75\mu$Ci}
   \label{ROC2}
\end{subfigure}
\begin{subfigure}[b]{0.24\textwidth}
   \includegraphics[height=1.75in]{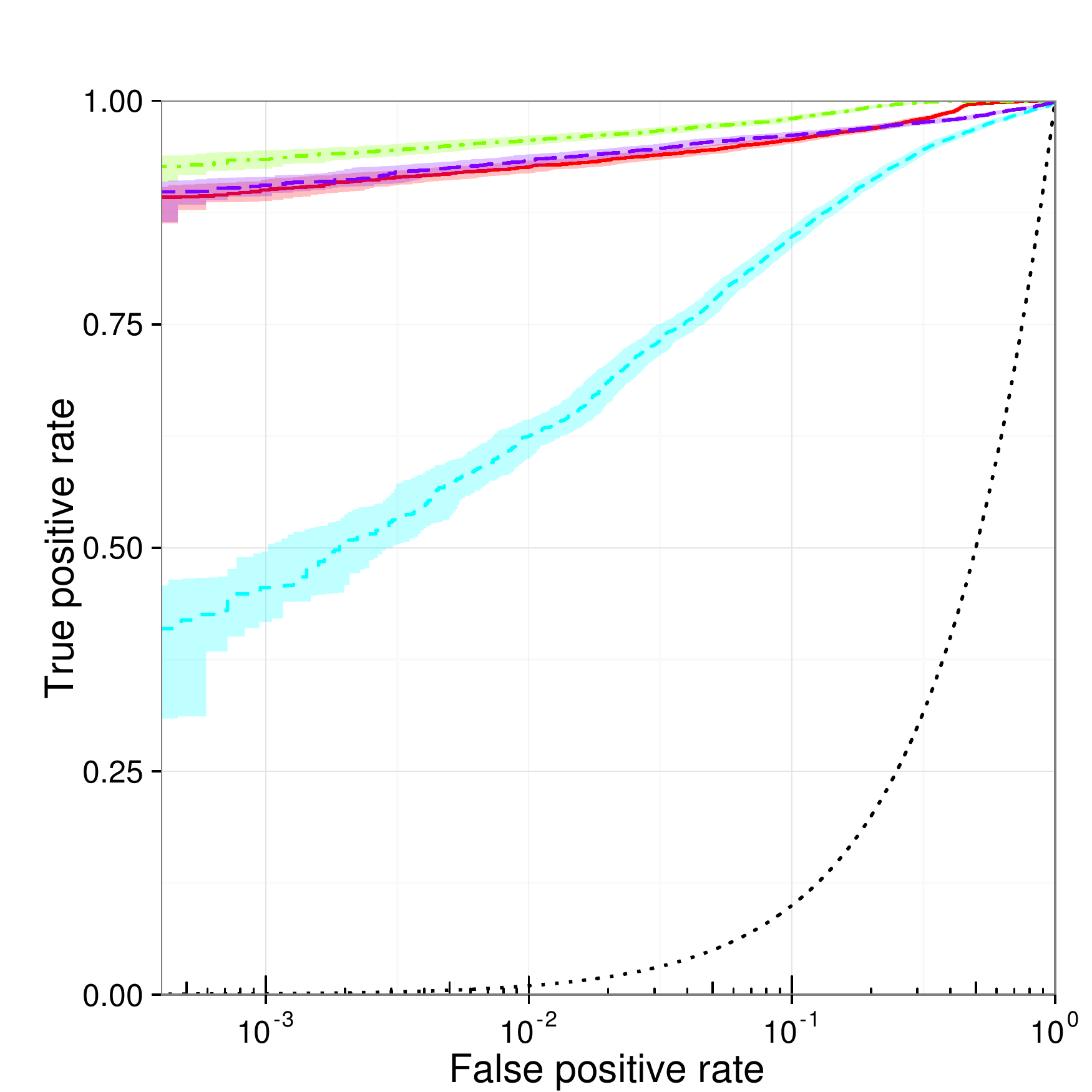}
   \caption{$100$-$250\mu$Ci}
   \label{ROC3}
\end{subfigure}
\begin{subfigure}[b]{0.1\textwidth}
\hspace*{-48pt}
\vspace*{20pt}
   \includegraphics[height=1.75in]{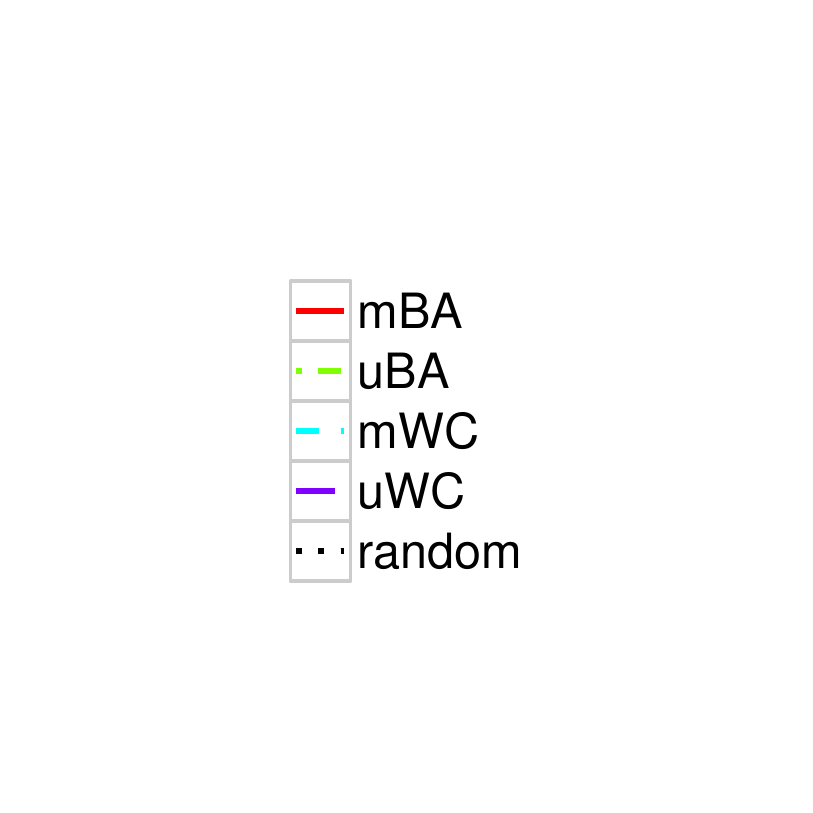}
\end{subfigure}
\caption{ROCs showing detection performance for each method and a selection of 3 source intensity bands.}
\label{ROCs}
\end{figure}

\begin{minipage}{\textwidth}
  \begin{minipage}[t]{0.33\textwidth}
    \centering
    \vspace*{10pt}
    \includegraphics[width=0.66\textwidth]{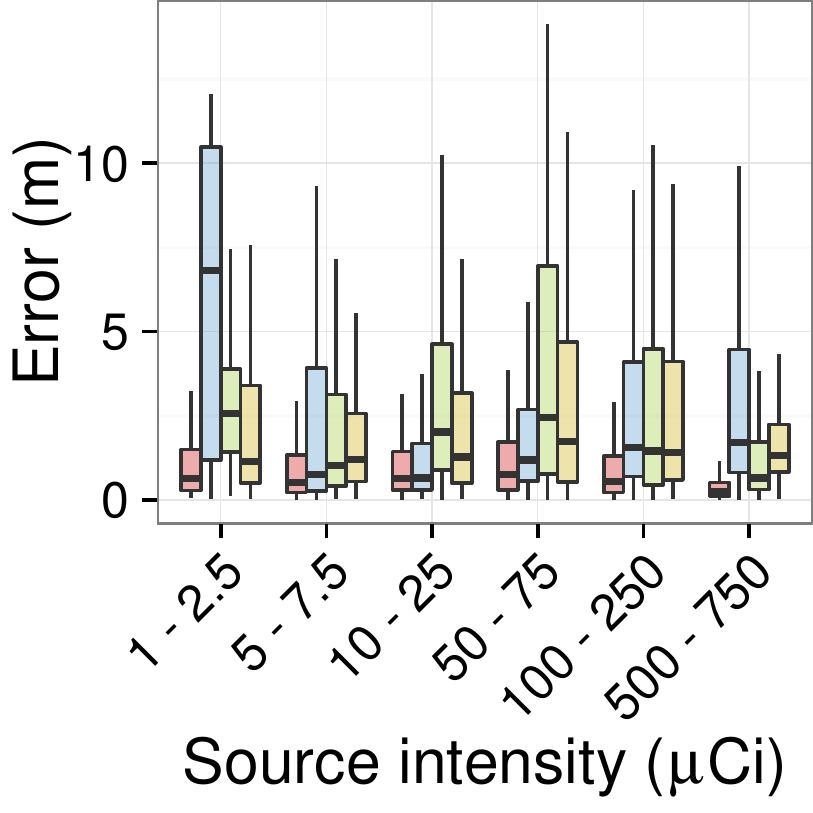}
    \includegraphics[trim={1in -1in 1in 0},clip,width=0.2\textwidth]{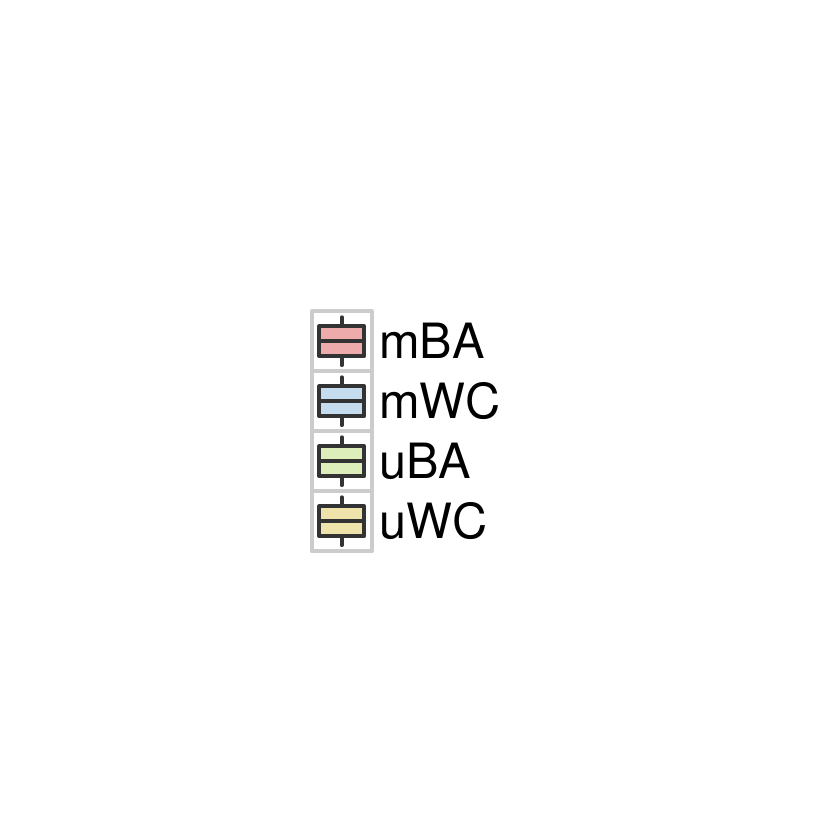}
    \label{BOXs}
    \vspace{-2pt}
    \captionof{figure}{Localization error for each band of source intensities and method.}
  \end{minipage}
  %\hfill
  \begin{minipage}[t]{0.67\textwidth}
    \centering
    \vspace{30pt}
    \captionof{table}{Probability of detection for each band of source intensities and method.}
    \begin{tabular}{c|cccccc}
    & \multicolumn{6}{c}{Source intensity ($\mu$Ci)} \\
   Method & 1-2.5 & 5-7.5 & 10-25 & 50-75 & 100-250 & 500-750 \\ \hline
   mBA & 0.011 & 0.077 & 0.207 & 0.631 & 0.902 & 0.971 \\
   uBA & 0.045 & 0.162 & 0.355 & 0.776 & 0.937 & 0.987 \\
   mWC & 0.003 & 0.024 & 0.070 & 0.243 & 0.457 & 0.889 \\
   uWC & 0.033 & 0.129 & 0.290 & 0.677 & 0.907 & 0.980
   \end{tabular}
  \end{minipage}
\end{minipage}

\vskip 0.05in
\noindent
{\bf Conclusions.}  From these experiments we can begin to draw boundaries around the problem space in which masked detectors provide positive utility. Our results suggest that problems with moderate source intensities and a high priority on localization may benefit from the use of masked detectors.

\vskip 0.05in
\noindent
{\bf References.} {[}1{]} TJ Aucott et al.\ Effects of Background on Gamma-Ray Detection for Mobile Spectroscopy and Imaging Systems. {\it IEEE Trans on Nuclear Science} (61)2, 986-991, 2014.

\vskip 0.05in
{\scriptsize
\noindent
This work has been partially supported by the U.S. Department of Energy under grant DE-NA0001736 and by the National Science Foundation under award 1320347.
Lawrence Livermore National Laboratory is operated by Lawrence Livermore National Security, LLC, for the U.S. Department of Energy, National Nuclear Security Administration under Contract DE-AC52-07NA27344
}

\end{document}